\pdfoutput=1
\documentclass[aps,prd,twocolumn,groupedaddress,nofootinbib,longbibliography]{revtex4-1}

\usepackage{amsfonts,amsmath,amssymb}
\usepackage{nicefrac}
\usepackage{graphicx}
\usepackage{color}
\usepackage{soul} 
\usepackage{hyperref}
\usepackage{accents}
\usepackage[normalem]{ulem}

\def\exd{\mathrm{d}}
\def\exD{\mathrm{D}}
\def\exi{\mathrm{i}}

\def\d{\delta}
\def\e{\epsilon}
\def\o{\omega}
\def\w{\wedge}

\def\k{\kappa}
\def\s{\sigma}

\def\cL{{\cal L}}
\def\cK{{\cal K}}
\def\L{\Lambda}
\def\cB{{\cal B}}
\def\Lambda{\lambda}
\def\Lz{L}

\def\half{\textstyle\frac12}

\def\beq{\begin{equation}}
\def\eeq{\end{equation}}
\def\bea{\begin{eqnarray}}
\def\eea{\end{eqnarray}}

\begin{document}


\title{Black hole entropy and Lorentz-diffeomorphism Noether charge}

\author{Ted Jacobson}
\email[]{jacobson@umd.edu}
\affiliation{Center for Fundamental Physics, University of Maryland, 
                    College Park, Maryland 20742, USA}
\author{Arif Mohd}
\email[]{amohd@umd.edu}
\affiliation{SISSA, 
Via Bonomea 265, 34136 Trieste, Italy \\
                INFN, Sezione di Trieste, Trieste, Italy}



\begin{abstract}
We show that, in the first or second order orthonormal frame formalism, black hole
entropy is the horizon Noether charge for a  combination of 
diffeomorphism and local Lorentz symmetry involving the Lie derivative of the frame. The Noether charge 
for diffeomorphisms alone is unsuitable, since a regular frame cannot 
be invariant under the flow of the Killing field at the bifurcation surface.
 We apply this formalism to 
 Lagrangians polynomial in wedge products of the frame field 1-form and curvature 2-form,
including general relativity, Lovelock gravity, and ``topological" terms in four dimensions.
\end{abstract}


\maketitle

\section{Introduction}

The entropy of black holes in any diffeomorphism invariant gravity theory
can be identified via a variational identity known as the first law of black hole
mechanics. In the approach of Wald \cite{Wald:1993nt}, this identity arises from considering
the Hamiltonian $H_\xi$ that generates evolution with respect to the flow of
the horizon-generating Killing vector $\xi$ of a black hole solution.
The variation $\delta H_\xi$ at a solution
is equal to a variation of boundary
terms, and vanishes because $\xi$ generates
a symmetry of the dynamical fields. When the boundaries lie at
the horizon bifurcation surface and at spatial infinity,
the implied relation between the boundary term variations is the first law,
from which Wald's formula for the black hole
entropy as Noether charge can be inferred.

This method is usually applied in a context where
the spacetime geometry is characterized by the metric tensor alone, however in some
settings it is necessary or desirable to use instead a 
formalism with geometry determined by an orthonormal frame and either the associated
spin connection (second order formalism) or an independent spin connection (first order formalism).
Application of Wald's method in this setting
appears at first to yield a vanishing
Noether charge at the bifurcation surface where $\xi$ 
vanishes --- and therefore vanishing black hole entropy --- because the Noether charge form 
involves $\xi$ without derivatives. The puzzle this raises has not to our knowledge
been discussed explicitly in the literature.

We trace the trouble to the
requirement that the frame (hereafter the `orthonormal' qualifier is implicit)
has vanishing
Lie derivative with respect to $\xi$.  This requirement cannot be met
at the bifurcation surface, and implies that the
derivative of the frame diverges at the bifurcation surface, so
that the spin connection diverges. 
On the other hand, the diffeomorphism Noether charge form
involves the contraction of the vanishing Killing vector with the diverging spin connection.
We first show how one can evaluate a finite, non-zero entropy by taking the
limit as the bifurcation surface is approached.

Next, in a second approach, we modify the derivation so that
the singular behavior does not arise in the first place. In a frame formalism the theory
is symmetric under both diffeomorphisms and local Lorentz
transformations of the frame. We show in this paper how the black hole entropy
can be derived as the Noether charge for a particular combination
of these symmetries. 
The frame can be invariant under the combined
symmetry associated with $\xi$, without having singular derivative at the horizon, so that
the extraction of the black hole entropy requires no limit.
The variation corresponding to this symmetry is defined by a ``Lorentz-Lie" 
derivative which is covariant under local Lorentz transformations of the frame field.
It is defined by adding to the ordinary Lie derivative a connection term built from the frame field and 
its partial derivatives. Besides allowing for nonsingular invariant frames at the bifurcation surface, 
this notion of combined Lorentz-diffeomorphism symmetry should allow the symmetry 
to be implemented on nonparallelizable manifolds, where no global frame field exists.
More generally, for theories containing fields charged under a gauge group $G$, the Noether charge formalism for symmetry 
under combined diffeomorphisms and local gauge transformations has been formulated recently in terms of fields living on a principal $G$-bundle
over spacetime \cite{Prabhu}.

This paper is organized as follows. Section \ref{Wald's algorithm} reviews the derivation showing that black-hole entropy is the horizon 
Noether charge associated with the diffeomorphism generated by the horizon-generating Killing vector field. 
In Section \ref{grwframes} we examine this Noether charge for general relativity in the frame formalism, 
diagnose the pathology, and treat it with a limit.  In Section \ref{LLderivative} we 
introduce the Lorentz-covariant Lie derivative, and in Section \ref{LLNoether} we show
how black hole entropy is the horizon Noether charge associated with the combined 
Lorentz and diffeomorphism symmetry it generates. 
In section \ref{sec:lovelock} we use this formalism to evaluate the black hole entropy in Lovelock theory in arbitrary dimensions, and in section \ref{topoterms} we 
apply it in four dimensions to evaluate the contributions of the Holst \cite{Holst:1995pc}, Euler and Pontryagin terms. We conclude in section \ref{sec:discussion} with a brief discussion.

We work in the units such that $16 \pi G = c=1$.  
 Lower case Greek letters are used for the spacetime indices,
 and internal Lorentz indices are denoted by lower case Latin letters. 
The metric signature is $({-}{+}{+}{+})$. 

\section{Black hole entropy as diffeomorphism Noether charge}
\label{Wald's algorithm}

In this section we sketch Wald's derivation \cite{Wald:1993nt} establishing that
black hole entropy is the diffeomorphism Noether charge
for the horizon-generating Killing field, evaluated
at the bifurcation surface. This will set the stage for the
application to the frame formalism, and our
modified derivation using a Lorentz-diffeomorphism
Noether charge.

Wald's derivation applies to any diffeomorphism invariant theory
defined by a Lagrangian $n$-form $L$, where $n$ is the spacetime dimension.
Denoting the dynamical fields collectively by $\phi$,
the variation $\d L$ induced by a field variation $\d \phi$ can be written as
\begin{align}
\label{deltaL}
\delta L = E\, \delta \phi + \exd \theta (\phi, \delta \phi).
\end{align}
The quantity $E$ defines the field equations, $E=0$. The $(n-1)$-form $\theta$
is constructed locally out of the dynamical fields and their first variation, and is
called the ``symplectic potential". The
anti-symmetrized field variation of $\theta$
defines an $(n-1)$-form, called the ``symplectic current",  via
\begin{align}
\label{Omega}
\Omega (\phi,\delta_1 \phi, \delta_2 \phi)=\delta_1 \theta(\phi,\delta_2 \phi) - \delta_2 \theta(\phi,\delta_1 \phi).
\end{align}
When integrated over a spatial initial value surface, $\Omega$ defines the
symplectic form on the phase space of solutions.

Now consider the variation induced by a diffeomorphism generated by
a vector field $\xi$,
\beq\label{deltaxiphi}
\delta_\xi \phi = \mathcal{L}_\xi \phi.
\eeq
Diffeomorphism invariance of the theory
means that the Lagrangian is constructed only from the dynamical fields, without
any background structure. In this case, the variation of $L$ induced by the
field variation $\delta_\xi \phi $ is equal to the Lie derivative of the Lagrangian itself,
\beq\label{deltaxiL}
\delta_\xi L = \mathcal{L}_\xi L = \exd \, \exi_\xi L.
\eeq
Since this is a total derivative we learn that the vector fields on the spacetime generate 
symmetries of the dynamics.
With each $\xi$ is associated an $(n-1)$-form called the Noether current form, defined as
\begin{align}
\label{j_xi}
j_\xi=\theta (\phi, \mathcal{L}_\xi \phi) - \exi_\xi L,
\end{align}
whose exterior derivative is given 
[according to \eqref{deltaL}, \eqref{deltaxiphi}, and \eqref{deltaxiL}] 
by
\begin{align}
\exd j_\xi = -E \mathcal{L}_\xi \phi.
\end{align}
For all vector fields $\xi$, the current $j_\xi$ is therefore closed ``on shell", i.e.\ when $E=0$.
This implies \cite{1990JMP....31.2378W} that, on shell, $j_\xi$ is an exact form,
\beq
\label{j=dQ}
j_\xi=\exd Q_\xi,
\eeq
where $Q_\xi$ is some $(n-2)$-form
that is constructed locally from the fields and their
derivatives.
 The integral of $Q_\xi$ over a closed $(n-2)$-surface $S$  is called the
 ``Noether charge" of $S$ relative to $\xi$ .

In the covariant framework used by Wald,
the space of solutions to the field equations is the
phase space of the theory, and the on shell variation $\d_\xi\phi$ is
the phase space flow vector corresponding to the 1-parameter family
of diffeomorphisms generated by $\xi$.
The Hamiltonian $H_\xi$ generating this flow
is related to the symplectic
form via Hamilton's equations,
$\delta H_\xi = \int_\Sigma \Omega (\phi, \delta \phi, \mathcal{L}_\xi \phi )$,
where $\Sigma$ is a Cauchy surface. On shell this variation
is a boundary term:
\begin{align}
\label{deltaH}
\delta H_\xi &= \int_\Sigma \Omega (\phi, \delta \phi, \mathcal{L}_\xi \phi)\\
&=  \int_\Sigma\delta \theta (\phi, \mathcal{L}_\xi \phi) - \mathcal{L}_\xi \theta(\phi, \delta \phi)\\
&=  \int_\Sigma \d j_\xi + \d (i_\xi L) - \exi_\xi \exd \theta -\exd \,\exi_\xi \theta\\
&= \oint_{\partial \Sigma} \delta Q_\xi - \exi_\xi \theta.\label{bt}
\end{align}
In the second line we used \eqref{Omega}, in the third line \eqref{j_xi}, and in
the fourth line \eqref{j=dQ} and \eqref{deltaL}.
If $\xi$ generates a symmetry of the fields in a solution $\phi$, then $\mathcal{L}_\xi \phi=0$,
and thus \eqref{deltaH} implies 
 $\d H_\xi=0$, so that \eqref{bt} yields an identity relating the surface term
variations away from that solution, $\oint_{\partial \Sigma}\delta Q_\xi - \exi_\xi \theta=0$.

Now consider a stationary, axisymmetric black hole with a Killing field $\xi$ that
generates a Killing horizon with nonzero, constant surface gravity $\kappa$,
and vanishes on a bifurcation surface $\cB$.
If we choose the hypersurface $\Sigma$ to have its only boundaries
at spatial infinity and at $\cB$, then the variational identity takes the form
\begin{align}
\label{deltaQ}
\oint_{\cB} \d Q_\xi = \oint_{\infty} \delta Q_\xi - \exi_\xi \theta,
\end{align}
where the orientations of both surfaces are induced by a vector pointing toward infinity.
The right hand side can be shown to be equal to  
$\delta \mathcal{E}-\Omega_H\, \delta \mathcal{J}$ where $\mathcal{E}$ and $\mathcal{J}$
are the asymptotically defined total energy and angular momentum, respectively,
and $\Omega_H$ is the angular velocity of the horizon. 
To evaluate the left-hand side, note that
since $\xi$ is a Killing vector, its
second and higher derivatives can be written in terms of $\xi$ and its
first derivative, together with the Riemann tensor and its derivatives,
so $Q_\xi$ depends on $\xi$ only algebraically via $\xi$ and $\nabla\xi$.
At $\cB$ the vector  $\xi$ vanishes, and 
\beq\label{n}
\nabla_\mu \xi^\nu = \partial_\mu \xi^\nu = \kappa n_\mu{}^\nu,
\eeq
where $n_{\mu\nu}$
is the binormal to $\cB$ (i.e.\ the normal 2-form, normalized to $-2$), oriented
as determined by the derivative of the Killing vector in \eqref{n}.
Hence all the $\xi$ dependence of $Q$ for the background solution
is contained in the specification of the
bifurcation surface and the (constant) surface gravity $\kappa$. 
Moreover, the replacement $\nabla_\mu \xi^\nu \rightarrow \kappa n_\mu{}^\nu$
may be made before the variation is taken:  
the quantity $a^\mu b_\nu\, \delta n_{\mu}{}^{\nu} $ vanishes unless $a^\mu$
is normal and $b^\nu$ is tangent to $\cB$, yet there are 
no normal-tangential components in the background tensor 
because they would not be invariant under 
the Killing flow of $\xi$ at $\cB$ (which acts as a boost normal to $\cB$).
The identity \eqref{deltaQ} therefore takes the form of the so-called first law of black hole 
thermodynamics
\beq
\label{1stlaw}
T_H\, \d S = \delta \mathcal{E}-\Omega_H \delta \mathcal{J},
\eeq
where $T_H={\hbar \kappa}/{2\pi}$ is the Hawking temperature,
and 
\beq
\label{S}
S = \frac{2\pi}{\hbar}\oint_\cB \widehat Q_\xi,
\eeq
where $\widehat Q_\xi$ (for the background as well as for the varied solution) 
is obtained from $Q_\xi$ by the replacement $\nabla_\mu \xi_\nu \rightarrow n_{\mu\nu}$.
The black-hole entropy $S$ is thus proportional to the horizon Noether charge
corresponding to the horizon-generating diffeomorphism. (For a more 
complete discussion see \cite{Iyer:1994ys}.) 

In order for the entropy to be nonzero, it would seem that
$Q_\xi$ must depend on $\nabla\xi$,
so $j_\xi$, and therefore $\theta (\phi, \mathcal{L}_\xi \phi)$,
must depend on $\nabla\nabla\xi$. Since the
Lie derivative of a tensor field depends on $\nabla \xi$, this requires that
$\theta(\phi,\d \phi)$ depends on at least one derivative of $\d\phi$,
and therefore that $L$ involves at least second derivatives.
Since the first order orthonormal frame formalism
involves only one derivative, it thus appears that the black hole entropy would
\textit{vanish} in that formalism, but that conclusion is obviously erroneous.  The
right hand side of the first law \eqref{1stlaw} is of course independent of
which formalism is used.
In the next section we compute the horizon Noether charge for general relativity
using the frame formalism, diagnose the flaw in the above reasoning,
and show how to evade the problem.

\section{Diffeomorphism Noether charge for general relativity with orthonormal frames}
\label{grwframes}

In the first-order orthonormal frame formalism, 
the Lagrangian for General relativity in n dimensions
is written in terms of the 
frame field 1-form $e^a$,  which is $SO(n-1,1)$  vector-valued, and the $SO(n-1,1)$ 
connection 1-form 
$\omega^{a}{}_b$. These are the independent
dynamical variables of the theory.  The spacetime
metric is given by $g_{\mu\nu}=\eta_{ab} e^a_\mu e^b_\nu$,
where $\eta_{ab}$ is the Minkowski metric, and the curvature 2-form 
is defined by $R^a{}_b=\exd \omega^a{}_b +\omega^a{}_c \wedge \omega^c{}_b$.
We raise and lower Lorentz indices with $\eta_{ab}$ and its inverse, $\eta^{ab}$.
We sometimes omit the Lorentz indices when that will not cause confusion.

The Lagrangian $n$-form for General Relativity in $n$-dimensions 
is a function of the frame and the spin-connection via the curvature 2-form, 
\beq\label{LGR}
L(e,\o)= \epsilon_{a\dots bcd}\, e^a\w\dots\w e^b\w R^{cd}.
\eeq
This is manifestly gauge invariant and diffeomorphism covariant.
The variation is given by 
\bea
\d L &=& \d e^a \w \frac{\partial L}{\partial e^a} +  \exD\d\o^{ab} \w \frac{\partial L}{\partial R^{ab}}\\
&=& \d e^a \w \frac{\partial L}{\partial e^a} +  \d\o^{ab} \w \exD\frac{\partial L}{\partial R^{ab}}\nonumber\\
&&~~~~~~~~~~~~~+ \exd\left(\d\o^{ab} \w \frac{\partial L}{\partial R^{ab}}\right).\label{readoff}
\eea
where $\exD$ is the Lorentz covariant exterior-derivative 
\cite{Cecile}, and we have used the identity $\delta R^{ab}= \exD \delta \omega^{ab}$.
(The variation forms are placed in the first position in order to avoid the need for a dimension-dependent minus sign that would arise when integrating by parts on the $\exD$.) 
The equations of motion are given by
\bea
\e_{abc\dots df}\, e^c\w\dots\w e^d\w\exD e^f &=& 0  ,\label{DB=0}\\
\epsilon_{ab\dots cde}\,e^b\w\dots \w e^c \wedge R^{de}  &=& 0.\label{ewR=0}
\eea
 The first of these equations implies (assuming $e^a$ is non-degenerate) the
 ``torsion-free" condition $\exD e^a=0$, which can be solved for the connection $\o=\o^e$.
When this is substituted in the second equation of motion, that becomes equivalent
to the vanishing of the Ricci tensor of $g_{\mu\nu}$, so one recovers the (vacuum) Einstein equation.
If one puts $\o=\o^e$ in the Lagrangian at the beginning, one has the second order frame 
formalism, and \eqref{DB=0} is true as an identity.
The diffeomorphism Noether current \eqref{j_xi} 
involves the Lie derivative of the connection, $\mathcal{L}_\xi\o$, which is given by
\beq\label{Lieomega}
\mathcal{L}_\xi \o =\exi_\xi \exd\o + \exd(\exi_\xi\o) = \exi_\xi R + \exD(\exi_\xi\o).
\eeq
Here we are treating the connection components as a collection of 1-forms, and we shall 
do the same with the frame components. If the relevant manifold cannot be covered by a single frame 
field --- i.e.\ is not parallelizable --- this strategy would not be available, because under a change of local Lorentz 
gauge the Lie derivative would not transform properly so as determine a well-defined symmetry operation. 
In that case, something like the Lorentz-Lie derivative discussed below would be required.

From \eqref{readoff} we can read off the symplectic potential defined in \eqref{deltaL},
\beq
\theta  = \d\o^{ab} \w \frac{\partial L}{\partial R^{ab}}.
\eeq
Using \eqref{Lieomega}, the diffeomorphism Noether current \eqref{jxi} can thus 
be written as
\begin{align}
\label{jxi}
&j_\xi = \exd\left(\exi_\xi\o^{ab} \w \frac{\partial L}{\partial R^{ab}}\right)\nonumber\\ 
&- (\exi_\xi\o^{ab})\w \exD \frac{\partial L}{\partial R^{ab}}
+ (\exi_\xi R^{ab}) \w \frac{\partial L}{\partial R^{ab}} -\exi_\xi L.
\end{align}
(In the first and second terms, the first factor is a 0-form, so the wedge product is just ordinary
multiplication. Throughout this paper we sometimes include such unnecessary wedge notations
since they seem helpful in organizing the structure of the expressions.)

The second term in the Noether current \eqref{jxi} vanishes by the $\o$ equation of motion. 
Moreover, the Lagrangian \eqref{LGR} has the nice property
\beq\label{nice}
\exi_\xi L = (\exi_\xi e^a) \w \frac{\partial L}{\partial e^a} +  (\exi_\xi R^{ab}) \w \frac{\partial L}{\partial R^{ab}},
\eeq
from which it follows that, taken together, the third and fourth terms of \eqref{jxi}
vanish by the $e$ equation of motion.
Thus we may simply read off the Noether charge $(n-2)$-form,
\beq\label{Noether charge}
Q_\xi = \exi_\xi\o^{ab} \w \frac{\partial L}{\partial R^{ab}}.
\eeq
Notice that this is linear in $\xi$, with no derivative on $\xi$.
If $\xi$ is a horizon generating Killing field, $Q_\xi$ therefore
appears to \textit{vanish} when evaluated at the
bifurcation surface $\cB$ of the Killing horizon. This would
imply that the entropy \eqref{S} vanishes,
but obviously something is wrong with this argument. \par

The problem arises because, in showing that the entropy is proportional to the horizon Noether charge,
we assumed that the dynamical fields have vanishing Lie derivative with respect to $\xi$.
Because of this, the connection $\o^e$ diverges as $\cB$ is approached.
We shall explain shortly from a geometric viewpoint \text{why} the connection diverges,
but first let us show that $i_\xi\o^e$ has a finite, nonzero limit at $\cB$, and use this to find the entropy.
 
The Lie derivative of the frame is given by
\begin{align}
\mathcal{L}_\xi e^a &= \exi_\xi \exd e^a + \exd \exi_\xi e^a\nonumber\\
&= \exi_\xi \exD e^a + \exD \exi_\xi e^a - \exi_\xi {\omega^a}_b \wedge e^b.
\end{align}
Setting this equal to zero, and using the field equation $\exD e^a=0$ (or the definition
of $\o^e$ in the second order formulation), we obtain
\begin{align}
\label{diverging connection}
\exi_\xi (\o^e)^a{}_b = e^\mu_b \exD_\mu (\exi_\xi e^a),
\end{align}
where $e^\mu_b$ is the inverse frame. To evaluate the 
right hand side note that 
the action of $\exD_\mu$ on tensors has not so far 
been specified (other than being torsion-free)
hence we may choose it to act on tensor indices as the torsion-free covariant
derivative $\nabla_\mu$ determined by the metric. 
With this choice we have ${\cal D}_\mu e_\nu^a = 0$, where ${\cal D}_\mu$ denotes
the \textit{full} derivative including both the spacetime and spin connections.
Then, using the Leibniz rule,
\eqref{diverging connection} becomes
\begin{align}
\label{diverging connection2}
\exi_\xi (\o^e)^a{}_b = e^\mu_b  e_\nu^a \nabla_\mu \xi^\nu.
\end{align}
The limiting value at $\cB$ is given by
\begin{align}
\label{diverging connectionatB}
\lim_{\rightarrow \cB} \exi_\xi (\o^e)^{ab} = -\kappa n^{ab},
\end{align}
where again $\kappa$ \eqref{n}
is the surface gravity, and $n^{ab}=n^{\mu\nu}e_\mu^a e_\nu^b$
is the bi-normal to $\cB$, converted to a Lorentz tensor. 
Thus, despite appearances, $\exi_\xi \o^e$ does \textit{not} vanish at
the bifurcation surface. This can only happen because $\o^e$ blows up there.

Using \eqref{diverging connectionatB}, we find the Noether charge 
form \eqref{Noether charge} is given by
\begin{align}
\label{Noether charge at B}
\lim_{\rightarrow \cB} (Q_\xi) = -\k n^{ab} \e_{abc\dots d}e^c\w\dots\w e^d.
\end{align}
This is just $2\k$ times the ``area" [$(n-2)$-volume] element on $\cB$, hence  
$\oint_\cB Q_\xi = \kappa A/8\pi G$ (restoring the $16\pi G$),
so the entropy \eqref{S} is $S_{BH} = A/4\hbar G$, the Bekenstein-Hawking entropy.

To explain why and how the connection diverges at the bifurcation surface, we employ
a simple analogy with a two dimensional Euclidean space.
The Killing vector field that generates the rotation around the origin is given by
$\xi=\partial_\theta$ in polar
coordinates $(r,\theta)$.
The origin is a fixed point of the rotational
isometry, i.e.\ $\xi$ vanishes there, so it is analogous to the bifurcation
 surface. A frame that has zero Lie derivative with respect to this rotation Killing field
 rotates by $2\pi$ when traversing a circle around the origin.
For a circle closer to the origin, the frame rotates
 faster, because the circumference shrinks. At the
 origin the frame has to rotate infinitely fast, which implies that the connection diverges.
 Explicitly, let the frame be
  given by $e^1= \exd r$ and $e^2 = r \exd \theta$, so that $\cL_\xi e^a=0$.
  The non-zero connection components are given by $\o^2{}_1=-\o^1{}_2 = \exd \theta$.
  The norm of $\exd \theta$ is $(g^{\theta\theta})^{1/2}=1/r$, so $\exd \theta$, and therefore the connection,
  diverges at the origin, although the contraction $\exi_\xi\o^2{}_1=1$ is finite and nonzero.
At the bifurcation surface of a black hole space-time one has a hyperbolic version
of this phenomenon.
For instance, for a Schwarzschild black hole we have $e^0=N \exd t$ and $e^1 = N^{-1} \exd r$, with $N=(1-2M/r)^{1/2}$ the norm of the Killing vector $\partial_t$. Then $\o^0{}_1=\kappa \exd t$, where $\k=1/4M$ is the surface gravity. The connection diverges since the norm of $\exd t$ is $N^{-1}$, although $\exi_{\partial_t}\o^0{}_1=\k$ is finite. 


If we are to avoid the occurrence of a singular spin connection in the Noether charge computation of
black hole entropy, we must modify the realization of the diffeomorphism symmetry, so that a frame can
be invariant under the symmetry and yet nonsingular at the bifurcation surface. The next section introduces this realization.

\section{Lorentz-Lie derivative}
\label{LLderivative}

The Lie derivative of tensor fields with respect to a vector field $\xi$ is defined,
with no additional structure, as the rate of change of the pull-back along the flow of $\xi$. 
A frame consists of co-vectors which are carried by the flow in a unique way.
The covectors  remain orthonormal under the flow of a Killing vector, but they undergo
a Lorentz transformation. Therefore 
the Lie derivative of a frame with respect to a Killing vector is generally nonzero. 
However, given a frame, one can define a modified derivative which includes a compensating 
local Lorentz transformation, so that the modified derivative of the frame with respect to 
a Killing vector is always zero. We call this the \textit{Lorentz-Lie (LL)  derivative}.
The LL derivative we employ has been introduced several times, using various formalisms
(see \cite{Kosmann1971,Henneaux1977f,Henneaux1977m,Jackiw:1979ub,PhysRevD.74.064002, Fatibene2011}
and references therein). 
Acting on a spinor field, the LL derivative agrees with the definition given by Kosmann \cite{Kosmann1971}. 
It was called the Yano derivative in \cite{PhysRevD.74.064002}, where other notions of 
generalized Lie derivative are also discussed.

We denote the Lorentz-Lie derivative by ${\mathcal K}^e_\xi$ (the notation is chosen in honor of Kosmann).
It is the Lie derivative supplemented
with a local $SO(n-1,1)$ gauge transformation generated by a particular $\Lambda^{e}_\xi$
which is determined by a frame $e^a$ as follows. Note first that metric compatibility, i.e.\
the vanishing of ${\mathcal K}^e_\xi\eta^{ab}$,  implies antisymmetry of 
$\Lambda^{e}_\xi$, that is, $(\Lambda^e_\xi)^{(ac)}=(\Lambda^e_\xi)^{(a}{}_b\eta^{c)b}=0$.
Now consider the action of ${\mathcal K}^e_\xi$ on $e^a$,
\beq
{\mathcal K}^e_\xi e^a ={\cal{L}}_\xi e^a + (\Lambda^e_\xi)^a{}_b e^b.  \label{Ke}
\eeq
The spacetime tensor $e_{a} {\mathcal K}^e_\xi e^a$ can be decomposed into its symmetric and anti-symmetric parts,
 \beq
 e_{a\mu} {\mathcal K}^e_\xi e^a_{\nu}=e_{a(\mu} {\mathcal K}^e_\xi e^a_{\nu)}+ e_{a[\mu} {\mathcal K}^e_\xi e^a_{\nu]}.\\
 \eeq
Owing to the antisymmetry of $(\Lambda^e_\xi)^{ab}$,  the symmetric part
is independent of $\Lambda^e_\xi$, and is given by
\begin{align}
e_{a(\mu} {\mathcal K}^e_\xi e^a_{\nu)}= \frac{1}{2} {\cal{L}}_\xi g_{\mu\nu}.
\end{align}
The LL derivative  
${\mathcal K}^e_\xi e^a$ will therefore vanish when $\xi$ is a Killing vector if and 
only if the antisymmetric part vanishes. 
The antisymmetric part, 
\begin{align}
e_{a[\mu} {\mathcal K}^e_\xi e^a_{\nu]}=e_{a[\mu} {\cal{L}}_\xi e^a_{\nu]} + e_{a\mu}e_{b\nu}(\Lambda^e_\xi)^{ab},
\end{align}
can be set to zero by choosing
\begin{align}
\label{Lambda}
(\Lambda^e_\xi)^{ab} = e^{\s[a} {\cal{L}}_\xi e^{b]}_{\s}.
\end{align}
This choice of $\Lambda^e_\xi$ defines the LL derivative associated with $e^a$.
The LL derivative of $e^a$ with respect to an arbitrary vector field  is thus given by
\begin{align}
{\mathcal K}^e_\xi e^a_\mu = \frac{1}{2} e^{a\nu} {\cal{L}}_\xi g_{\mu\nu}.
\end{align}
In particular, when $\xi^a $ is a Killing vector field we have ${\mathcal K}_\xi e^a = 0$.

It will be useful to find an explicit expression for $\lambda^e_\xi$ (\ref{Lambda}) 
in terms of $\nabla\xi$.
We have
\begin{align}
\label{Lambda-explicit1}
(\Lambda^e_\xi)^{ab} &= e^{\mu[a} {\cal{L}}_\xi e^{b]}_{\mu}\\
&=e^{\mu[a}\xi^\nu \nabla_\nu e^{b]}_{\mu} + e^{\mu[a} (\nabla_\mu \xi^\nu) e^{b]}_{\nu}\label{Lambda-explicit2}\\
&= \exi_\xi (\o^e)^{ab} + e^{\mu [a} e_\nu^{b]}\nabla_\mu \xi^\nu.\label{Lambda-explicit3}
\end{align}
In the second line we expressed the Lie derivative using the torsion-free
metric compatible derivative $\nabla$, and in the third line we used 
$\nabla e^b = {\cal D} e^b - (\o^e)^b{}_c e^c = - (\o^e)^b{}_c e^c$.

Under a Lorentz transformation of the frame, $e^a\rightarrow \Lz^a{}_b e^b$, the quantity
$\lambda_\xi^e$ transforms like a connection for the Lie derivative, 
\beq
\lambda_\xi^{\Lz e} = \Lz \lambda_\xi^e \Lz^{-1} + \Lz {\mathcal L}_\xi \Lz^{-1}.
\eeq
This makes the LL derivative covariant under $SO(n-1,1)$ gauge transformations.
The action of the LL derivative is extended to any Lorentz tensor by requiring that
it be a derivation, i.e.\ by stipulating that the Leibniz product rule applies.
Its action on any $SO(n-1,1)$ connection is defined so that the $\lambda_\xi$ term 
implements the infinitesimal gauge transformation of a connection,
\begin{align}
\cK^e_\xi \o^{ab} &= \cL_\xi\o^{ab} - \exD(\L^e_\xi)^{ab} \\
&= \exi_\xi R^{ab} + \exD(\exi_\xi\o -\L^e_\xi)^{ab}\label{Kdiffomega2}
\end{align}
This result will be key when evaluating the entropy using 
the Lorentz-diffeomorphism Noether charge. 

Let us illustrate the action of the LL derivative in two-dimensional flat Euclidean space.
The frame we considered 
above
has zero Lie derivative along the rotation Killing vector field 
$\xi=\partial_\theta$.
Hence for that frame and that vector field we have $\Lambda_\xi=0$, so the LL derivative is just the Lie derivative,
which vanishes on the frame.
The problem with such a frame, as explained above, is that it is singular at the fixed point of the Killing flow. Next
we consider a Cartesian frame, $e^{1} = dx$ and  $e^{2}=dy$.
Writing the  same Killing vector as $\xi= x\, \partial_y-y\, \partial_x$,
it is simple to see that $(\cL_\xi e)^1 = -e^2$ and $(\cL_\xi e)^2 = e^1$. Although this frame is not rotationally
 invariant, its LL derivative must vanish since $\xi$ is a Killing field.
Indeed we have $(\L^e_\xi)^1{}_2 = -(\L^e_\xi)^2{}_1=1$, so $(\cK^e e)^1= (\cL_\xi e)^1 + (\L^e_\xi)^1{}_2\,e^2 = -e^2 + e^2 = 0$,
and similarly $(\cK^e e)^2 = (\cL_\xi e)^2 + (\L^e_\xi)^2{}_1\,e^1 = e^1 - e^1 = 0$.
In effect, the gauge transformation cancels the nonzero Lie-derivative with respect to a Killing vector.
(If we consider instead the shear vector field $x\,\partial_y$, which is \textit{not} a Killing vector,
then both the Lie and LL derivatives of the Cartesian frame are non-vanishing, and they differ from each other.) 
Similarly, the rotation invariant frame has non-vanishing
Lie derivative with respect to the translation Killing vector $\partial_x$, 
but its LL derivative with respect to $\partial_x$ vanishes. 

\section{Black hole entropy as Lorentz-diffeomorphism  Noether charge}
\label{LLNoether}

We may now repeat the steps in the Noether charge construction of Sec.~\ref{Wald's algorithm},
replacing the Lie derivative variation by the LL derivative,
\beq\label{Kphi}
\d \phi = \cK^e_\xi\phi.
\eeq
Assuming the diffeomorphism-covariant Lagrangian is a Lorentz scalar, its variation is the same 
whether the fields of which it is built vary by the Lie derivative, or the LL derivative, hence it satisfies
$\cK_\xi^e L=\cL_\xi L= \exd\,\exi_\xi L$. 

The Noether current associated with the LL symmetry is defined by 
\beq\label{jK}
j_\xi^\cK = \theta(\phi, \cK^e_\xi \phi) -\exi_\xi L,
\eeq
which is closed on shell for all $\xi$, and 
hence is the exterior
derivative of a Noether charge $(n-2)$-form,
\beq\label{QK}
j_\xi^\cK = \exd Q_\xi^\cK.
\eeq
The derivation of the first law of black hole mechanics proceeds as in the
case of the diffeomorphism Noether current, but the role of the Lie derivative 
is played by the LL derivative. In particular, to make use of the correspondingly 
modified variational identity \eqref{bt}, 
the background fields must now satisfy $\cK^e_\xi\phi=0$, so that the variation of the Hamiltonian 
generating the combined Lorentz-diffeomorphism symmetry will vanish.
This leads to a new expression
for the black hole entropy,
\beq
\label{SK}
S = \frac{2\pi}{\hbar}\oint_\cB \widehat Q^\cK_\xi,
\eeq
where again the hat on $Q$ indicates the replacement
$\nabla_\mu \xi_\nu \rightarrow n_{\mu\nu}$.
In order to evaluate this for a particular theory one needs first to find
the Noether current and then the corresponding Noether charge form.
Let us see how it works out for General Relativity and some closely related theories.

\subsection{Lovelock gravity}
\label{sec:lovelock}

The analysis for General Relativity in section \ref{grwframes} actually applies more generally to any Lagrangian
$L(e,\o)$ that is constructed from wedge products of frames and curvature 2-forms, since 
the nice property \eqref{nice} continues to hold, and the rest of the derivation is generic. 
In particular, the expression for the Noether charge form \eqref{Noether charge} applies to all such Lagrangians. 
These Lagrangians correspond to Lovelock gravity theories, together with various ``topological" terms that do not affect the equations of motion.

Comparison of the expressions (\ref{Kdiffomega2}) and (\ref{Lieomega}) for the LL and Lie derivatives
of the connection reveals that, to obtain the Noether charge form,
we merely need to replace $\exi_\xi\o$ by $\exi_\xi\o-\lambda_\xi^e$
in \eqref{Noether charge}. This yields
\beq\label{QKexplicit}
Q^\cK_\xi = (\exi_\xi\o -\L^e_\xi)^{ab} \w \frac{\partial L}{\partial R^{ab}}.
\eeq
The key point now is that since the frame is LL invariant and not Lie invariant,
it can be assumed to be regular at $\cB$. Therefore the quantity $\exi_\xi\o$ vanishes
at $\cB$, and from \eqref{Lambda-explicit3}
 \eqref{n} we have there
\beq\label{LatB}
(\L^e_\xi)^{ab} = \kappa n^{ab}.
\eeq
When this is substituted in \eqref{QKexplicit}, the result is identical to 
what we obtained using the limiting expression \eqref{diverging connectionatB} 
with a singular, Lie invariant frame. That is, 
\beq
\widehat Q^\cK_\xi = -\kappa n^{ab} \w \frac{\partial L}{\partial R^{ab}},
\eeq
and integrating this gives the entropy \eqref{SK}. 

The Lagrangian for Lovelock gravity is 
\begin{align}
L(e,\o)= \epsilon_{a\dots bcd}(&c_0\, e^a\w\dots\w e^b\w e^c\w e^d\nonumber\\ + &c_1\, e^a\w\dots\w e^b\w R^{cd}
+ \dots),\label{Love}
\end{align}
where $c_i$ is a coupling constant for the term with $i$ factors of the curvature, and the terms 
indicated by the ellipses each contain one more factor of $R$ than the previous term. 
The $c_0$ term is a cosmological constant, and the $c_1$ term is the Einstein-Hilbert Lagrangian. 
The form $Q^\cK_\xi$ is obtained from
$L$ by moving, in turn, each factor of $R$ all the way to the first position and replacing it by $-\kappa n^{ab}$. 
Contracting $n^{ab}$ with the rank-$n$ Lorentz $\epsilon$ in $L$ produces twice the 
rank-$(n-2)$ Lorentz $\epsilon$ associated via the frame with the $SO(n-2)$ group of the tangent space of 
$\cB$. The remaining  Lorentz indices are thus all projected into this subspace. 
The coefficient of the term in $Q^\cK_\xi$ with $m-1$ factors of $R$ is thus 
$2\k m c_m$.

The curvatures in the entropy integrand are those of 
the connection $\o$, 
whose equation of motion is 
$\exD\, \partial L/\partial R^{ab}=0$.
One way --- and generically  the only way --- to satisfy this is to have 
$\exD e^a=0$, i.e.\ for $\o$ to be the spin connection $\o^e$
determined by $e$. For such solutions the curvature appearing
in the entropy is the one determined by $e$. 
These curvature 2-forms are all pulled back to $\cB$ and, 
as explained above, their Lorentz indices are all projected 
into the $\cB$-subspace.
Moreover, the extrinsic curvature of the bifurcation surface
vanishes, so these curvatures all reduce to intrinsic curvatures of $\cB$. 
The entropy is therefore determined by the \textit{intrinsic} geometry of the horizon \cite{Jacobson:1993xs}.

The first and second order
formalisms for Lovelock gravity are not strictly identical in more than four dimensions, since 
there exist solutions in the first order formalism for 
which $\o\ne\o^e$. That is, the connection may 
have torsion. In fact, black hole solutions with this property 
exist, and their entropy might involve this torsion via the curvature 
(see for example \cite{Canfora:2010rh} and references therein).
In \cite{PhysRevD.84.084015} black hole solutions in Born-Infeld gravity  
(which is a special case of Lovelock gravity in even dimensions) 
supporting non-zero torsion were constructed. However, by construction all the Noether charges for these solutions vanish, including the entropy. 
It would be interesting to find solutions with non-trivial torsion contributing to the black hole entropy.

\subsection{``Topological" terms}
\label{topoterms}

As a further application of the Lorentz-diffeomorphism symmetry discussed here, we now 
look into the contributions of ``topological" terms to the black hole entropy in four dimensional general relativity.
The contributions of these terms have been studied before, using various formalisms;
 see for example Refs.~\cite{Corichi:2010ur,Corichi:2013zza,Fatibene2011,Durka:2011yv}.

The Lagrangian 4-form is given by 
\begin{align}
L(e,\o) &= \Bigl(*(e^a \w e^b)\nonumber\\  &+ c_{\rm H}\,  e^a\w e^b
+c_{\rm E}\,  *\!R^{ab}  + c_{\rm P}\,  R^{ab}\Bigr)\w R_{ab},
\label{Ltop}
\end{align}
where 
 denotes Lorentz dual, e.g.\  $*R^{ab}=\half\e^{abcd}R_{cd}$.
The coupling constants are $c_{\rm H}$ for the Holst term \cite{Holst:1995pc},
$c_{\rm E}$ for the Euler (Gauss-Bonnet) invariant, and  $c_{\rm P}$ for the Pontryagin invariant. 
The Holst term modifies the connection equation of motion, but does not
affect its solution $\o^e$, and it drops out of the frame equation of motion when
the connection is on shell. The Euler and Pontryagin terms depend only on the connection. 
The Euler and Pontryagin terms are exact forms, so do not affect the equations of motion. 
Were they exterior derivatives of gauge-invariant forms, we could absorb those 
forms into the symplectic potential $\theta$ \eqref{deltaL}, and from general considerations 
conclude that the entropy is unaffected by them \cite{PhysRevD.49.6587}. However, those forms are not
gauge invariant, hence these terms might contribute to the black hole entropy.

The black hole entropy (\ref{SK}) for the Lagrangian \eqref{Ltop} is given by 
\beq
S=\frac{2\pi}{\hbar}\oint_\cB n^{cd}\bigl(*e_c\w e_d + c_{\rm H}  e_c\w e_d 
+2c_{\rm E} *\! R_{cd} + 2c_{\rm P}  R_{cd}\bigr).
\eeq
The Einstein-Hilbert term is proportional to the area of $\cB$, as we saw before.
The Holst term vanishes because the binormal is orthogonal to $\cB$. 
The Euler term is one of the terms in the general Lovelock 
Lagrangian \eqref{Love}. Therefore, as explained above, it involves only
the intrinsic curvature of $\cB$. In the present case, since $\cB$ is two-dimensional, 
that just amounts to the Ricci scalar. The integral of this term in the 
entropy is a topological invariant, proportional to the Euler characteristic 
of the horizon \cite{Jacobson:1993xs}. In higher, even dimensions, similar terms
exist, involving $(n-2)/2$ curvature tensors.  
Finally, it turns out that, since the extrinsic curvature vanishes, 
the Pontryagin term is an exact form on $\cB$,
so its integral vanishes.
To see that 
the pull-back of $n^{cd}R_{cd}$ to $\cB$ is exact, let $l^a$ and $n^a$ be null
normals to $\cB$ satisfying $l_c n^c=-2$, so $n^{cd} = l^{[c} n^{d]}$. Then we have 
$n^{cd}R_{cd} = l_c D^2 n^c = d(l_c D n^c) - Dl_c\w Dn^c$.
Since the extrinsic curvature of $\cB$ vanishes, the null normals must 
be parallel transported along $\cB$ into multiples of themselves, so 
pulled back to $\cB$ we have $Dl^c = \s l^c$ and 
$Dn^c = -\s n^c$ for some 1-form $\s$. Hence $Dl_c\w Dn^c= -2 \s\w\s =0$.

\section{Discussion}
\label{sec:discussion}

In this paper we have made use of the Lorentz-Lie derivative ${\cal K}_\xi$ to define a particular variation of the frame field (and other Lorentz tensors) under a diffeomorphism generated by a vector field $\xi$. In words, the LL derivative  is defined by combining the usual Lie derivative with a term that subtracts the local Lorentz transformation induced on the frame by the flow. This subtraction term depends on the frame field, and amounts to a connection that covariantizes the Lie derivative with respect to local Lorentz transformations. A key property of this definition is that if $\xi$ is a Killing vector, the LL derivative of the frame vanishes. This property makes it possible for the frame to be LL-invariant at the bifurcation surface of a Killing horizon while remaining regular there.
Using this formalism, we showed how the LL Noether charge yields the black hole entropy. We illustrated the computational convenience of this method by evaluating the black hole entropy for Lagrangians that are polynomial in wedge products of the frame field 1-form and curvature 2-form.

The computations in this paper were carried out using a single ``local Lorentz gauge", so in effect we assumed that the relevant portion of the spacetime could be covered by a single gauge patch. Further analysis would be required to deal with situations where that is not the case. For example, one could use the frame bundle formalism, which has been discussed in this setting in Refs.~\cite{Fatibene2011,Prabhu}.

We have restricted attention here to Lagrangians that are Lorentz scalar $n$-forms. It could be interesting to study the Noether charge formalism allowing for Lagrangians having this property only up to the exterior derivative of a 
non-scalar $n$-form. This would shed a different light on the contributions of the Euler and Pontryagin terms studied here, and could be useful in further generalizations.

Finally, the combined diffeomorphism-gauge Noether current analysis can also be applied when the local gauge symmetry is internal, as in Yang-Mills theory. A simple example involving the electromagnetic field is discussed in Appendix E1 of \cite{Gralla:2014yja}. It employs the notion of ``gauge covariant Lie derivative" to arrive directly at a gauge-covariant Noether current. A general analysis is provided in Ref. \cite{Prabhu}.


\begin{acknowledgments}
We thank L. Bombelli for stimulating discussions,  L. Fatibene, and M. Henneaux for helpful
correspondence about their work, and K. Prabhu for comments on the manuscript.
The research of TJ was supported in part by Perimeter Institute for Theoretical Physics.  Research at Perimeter Institute is supported by the Government of Canada through Industry Canada and by the Province of Ontario through the Ministry of Research \& Innovation.  TJ was also supported in part by the NSF under grants number PHY-0903572, PHY-1407744, and PHY11-25915. The research of AM  was made possible through the support of a grant from the John Templeton Foundation. The opinions expressed in this publication are those of the authors  and do not necessarily reflect the views of the John Templeton Foundation.

\end{acknowledgments}

\bibliography{NoetherFramesBib}

\end{document}